\documentclass[aps,prl,twocolumn,showpacs,superscriptaddress]{revtex4-1}

\usepackage{amsfonts,mathrsfs,amsmath,amsthm,amssymb}
\usepackage{bm,times}
\usepackage{graphicx,epsfig}
\usepackage[colorlinks=true,breaklinks=true,linkcolor=blue,citecolor=blue,urlcolor=blue]{hyperref}

\def \non {\nonumber}

\newcommand{\be}{\begin{eqnarray}}
\newcommand{\ee}{\end{eqnarray}}

\makeatletter
    
    \newcommand{\Rmnum}[1]{\expandafter\@slowromancap\romannumeral #1@}
\makeatother

\begin{document}
\title{The advantage of the concatenated three-qubit codes}
\author{Long Huang}
\email{huangl@sicnu.edu.cn}
\affiliation{Coll Phys \& Elect Engn, Sichuan Normal University, Chengdu 610101, China}

\begin{abstract}
In this work, the efficient quantum error-correction protocol against the general independent noise is constructed with the three-qubit codes. The rules of concatenation are summarized according to the error-correcting capability of the codes. The codes not only play the role of correcting errors, but the role of polarizing the effective channel. For any independent noise, the most suitable error-correction protocol can be constructed based on the rules of concatenation. The most significant aspect of using the concatenated three-qubit codes is to realize quantum error-correction with the non-ideal gate operations, because the error-correction schemes of the three-qubit codes are simple and modular. For example, for the amplitude damping noise with the initial channel fidelity $0.9$, the effective channel fidelity can reach $0.94731$ (or $0.961634$ with the ideal quantum gate operations) when using the $4$ levels concatenated quantum error-correction with the three-qubit codes. The protocol with the three-qubit codes needs $81$ qubits and $366$ quantum gate operations when the accuracy rate is $\sqrt{0.999}$. Meanwhile, when using the $3$ levels concatenated quantum error-correction with the five-qubit code, the effective channel fidelity can only reach $0.922798$ (or $0.975488$ with the ideal quantum gate operations). The protocol with the five-qubit code needs $125$ qubits and $1147$ quantum gate operations when the accuracy rate is $\sqrt{0.999}$. The physical resources costed is multiple of the physical resources costed by realizing the three-qubit quantum error-correction, the protocol has a higher fault tolerance threshold, and no increase in complexity. So, we believe it will be helpful for realizing quantum error-correction in the physical system.
\end{abstract}

\date{\today}

\maketitle

\emph{Introduction.}---In quantum computation and communication, quantum error-correction (QEC) developed from classic schemes to preserving coherent states from noise and other unexpected interactions. It was independently discovered by Shor and Steane~\cite{Shor,Steane}. The QEC conditions were proved independently by Bennett, DiVincenzo, Smolin and Wootters~\cite{Bennett}, and by Knill and Laflamme~\cite{Knill}. QEC codes are introduced as active error-correction. The nine-qubit code was discovered by Shor, called the Shor code. The seven-qubit code was discovered by Steane, called the Steane code. The five-qubit code was discovered by Bennett, DiVincenzo, Smolin and Wootters~\cite{Bennett}, and independently by Laflamme, Miquel, Paz and Zurek~\cite{Laflamme}.

There are many constructions for specific classes of quantum codes. Rains, Hardin, Shor and Sloane~\cite{Rains} have constructed interesting examples of quantum codes lying outside the stabilizer codes. Gottesman~\cite{Gottesman2} and Rains~\cite{Rains2} construct non-binary codes and consider fault-tolerant computation with such codes. Aharonov and Ben-Or~\cite{Aharonov} construct non-binary codes using interesting techniques based on polynomials over finite fields, and also investigate fault-tolerant computation with such codes. Approximate QEC can lead to improved codes was shown by Leung, Nielsen, Chuang and Yamamoto~\cite{Leung}.

Calderbank and Shor~\cite{Calderbank}, and Steane~\cite{Steane2} used ideas from classical error-correction to develop the CSS (Calderbank-Shor-Steane) codes. Calderbank and Shor also stated and proved the Gilbert-Varshamov bound for CSS codes. Gottesman~\cite{Gottesman} invented the stabilizer formalism, and used it to define stabilizer codes, and investigated some of the properties of some specific codes. Independently, Calderbank, Rains, Shor and Sloane~\cite{Calderbank2} invented an essentially equivalent approach to QEC based on ideas from classical coding theory.

QEC codes are introduced as active error-correction. Another way, passive error-avoiding techniques contain decoherence-free subspaces~\cite{Duan,Lidar,Zanardi} and noiseless subsystem~\cite{KandLV,Zanardi2,Kempe}. Recently, it has been proven that both the active and passive QEC methods can be unified~\cite{Kribs,Poulin 05,Kribs2}. So, more QEC codes means more options for suppressing noise, and more options for optimizing the performance of QEC. Meanwhile, how to realize the efficient QEC in the physical system is still an important question because of the non-ideal quantum gate operations.

As known, the three-qubit code is not efficient for arbitrary one-qubit errors because of the Hamming bound. The three-qubit code is bound to cause channel fidelity to improve when the weight of the polarizing error (Pauli-X, Y, or Z) is above $\frac{2}{3}$. The three-qubit code in Eq.~(\ref{e1}) will change the polarizing of the effective channel to Pauli-Z after QEC, and the three-qubit code in Eq.~(\ref{e2}) will change the polarizing of the effective channel to Pauli-X after QEC. Based on the performances, under certain conditions, the QEC for arbitrary one-qubit errors can be realized when the two three-qubit codes are applied in concatenated.

\begin{figure}[tbph]
\centering
\includegraphics[width=0.4 \textwidth]{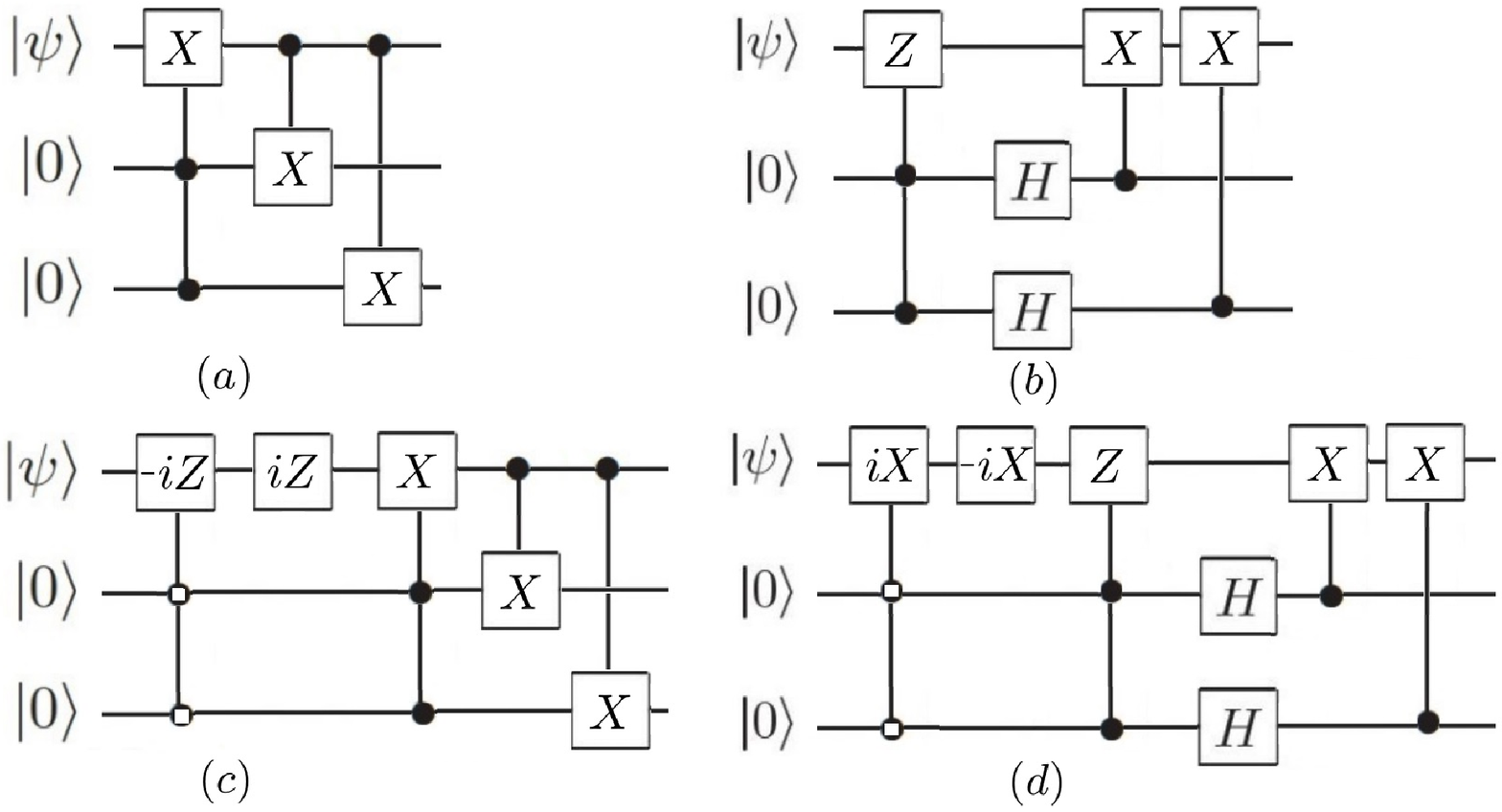}\\
\caption{(a) Encoding circuits for the bit-flip error.
 (b) Encoding circuits for the phase-flip error. (c) Encoding circuits for the bit-phase-flip error with the code in Eq.~(\ref{e1}).
 (d) Encoding circuits for the bit-phase-flip error with the code in Eq.~(\ref{e2}). }\label{figure1}
\end{figure}

On the other hand, the gate operations used in the QEC protocol are always none-ideal, which may result in the failure of the overall QEC scheme. Because the three-qubit codes are the smallest code, and we speculate that it may have a higher fault tolerance threshold with the none-ideal gate operations. Through calculation, we obtain the exact gate operations for the-qubit codes in the Fig.~(\ref{figure1}), and the fault tolerance threshold has an exponential improvement over the fault tolerance threshold of the five-qubit code in ref.~\cite{Bennett}. It will be helpful for realizing QEC in the physical system.

\emph{Three-qubit codes.}---In the first place, we introduce the two simplest three-qubit QEC codes in Eq.~(\ref{e1}) and~(\ref{e2}), because only the two simplest three-qubit codes can be used for encoding the physical qubits. One is the bit-flip code, which is the simplest code with the stabilizer $\langle Z_1Z_2, Z_2Z_3\rangle$,
\be
\label{e1}
|0_\mathcal{L}\rangle=|000\rangle,\non\\
|1_\mathcal{L}\rangle=|111\rangle.
\ee
With the code, the state $|\psi\rangle=a|0\rangle+b|1\rangle$ is encoded to $|\tilde{\psi}\rangle=a|000\rangle+b|111\rangle$. For the code in Eq.~(\ref{e1}), the correctable errors are $X_{1},X_{2},X_{3}$, $Z_{1}Z_{2},Z_{1}Z_{3},Z_{2}Z_{3}$, $Y_{1}Z_{2},Y_{1}Z_{3},Y_{2}Z_{3}$, $Z_{1}Y_{2},Z_{1}Y_{3},Z_{2}Y_{3}$, and $Z_{1}Z_{2}X_{3},Z_{1}Z_{3}X_{2},Z_{2}Z_{3}X_{1}$. The measurement-free QEC protocol is shown in Fig.~(\ref{figure1}a). (For Pauli-Y errors, the correctable errors are replacing the $X$ as $Y$, and the $Y$ as $X$). Because the list of correctable errors result in the same states as $X_{1},X_{2},X_{3}$, which are identifiable and correctable,
\be
X_1|\tilde{\psi}\rangle&=&a|100\rangle+b|011\rangle,\non\\
X_2|\tilde{\psi}\rangle&=&a|010\rangle+b|101\rangle,\non\\
X_3|\tilde{\psi}\rangle&=&a|001\rangle+b|110\rangle.\non
\ee
For Pauli-Y errors, the list of correctable errors result in the same states as $Y_{1},Y_{2},Y_{3}$, which are identifiable and correctable,
\be
Y_1|\tilde{\psi}\rangle&=&i[a|100\rangle-b|011\rangle],\non\\
Y_2|\tilde{\psi}\rangle&=&i[a|010\rangle-b|101\rangle],\non\\
Y_3|\tilde{\psi}\rangle&=&i[a|001\rangle-b|110\rangle].\non
\ee
The measurement-free QEC protocol is shown in Fig.~(\ref{figure1}c).

The other is the phase-flip code, which is the simplest code with the stabilizer $\langle X_1X_2, X_2X_3\rangle$,
\be
\label{e2}
|0_\mathcal{L}\rangle=\frac{1}{2}[|000\rangle+|011\rangle+|101\rangle+|110\rangle],\non\\
|1_\mathcal{L}\rangle=\frac{1}{2}[|111\rangle+|100\rangle+|010\rangle+|001\rangle].
\ee
With the code, the state $|\psi\rangle=a|0\rangle+b|1\rangle$ is encoded to $|\tilde{\psi}\rangle=a[|000\rangle+|011\rangle+|101\rangle+|110\rangle]+b[|111\rangle+|100\rangle+|010\rangle+|001\rangle]$. For the code in Eq.~(\ref{e2}), the correctable errors are $Z_{1},Z_{2},Z_{3}$, $X_{1}X_{2},X_{1}X_{3},X_{2}X_{3}$, $Y_{1}X_{2},Y_{1}X_{3},Y_{2}X_{3}$, $X_{1}Y_{2},X_{1}Y_{3},X_{2}Y_{3}$, and $X_{1}X_{2}Z_{3},X_{1}X_{3}Z_{2},X_{2}X_{3}Z_{1}$. The measurement-free QEC protocol is shown in Fig.~(\ref{figure1}b). (For Pauli-Y errors, the correctable errors are replacing the $Z$ as $Y$, and the $Y$ as $Z$). Because the list of correctable errors result in the same states as $Z_{1},Z_{2},Z_{3}$, which are identifiable and correctable,
\be
Z_1|\tilde{\psi}\rangle&=&a[|000\rangle+|011\rangle-|101\rangle-|110\rangle]\non\\
&&+b[-|111\rangle-|100\rangle+|010\rangle+|001\rangle],\non\\
Z_2|\tilde{\psi}\rangle&=&a[|000\rangle-|011\rangle+|101\rangle-|110\rangle]\non\\
&&+b[-|111\rangle+|100\rangle-|010\rangle+|001\rangle],\non\\
Z_3|\tilde{\psi}\rangle&=&a[|000\rangle-|011\rangle-|101\rangle+|110\rangle]\non\\
&&+b[-|111\rangle+|100\rangle+|010\rangle-|001\rangle].\non
\ee
For Pauli-Y errors, the list of correctable errors result in the same states as $Y_{1},Y_{2},Y_{3}$, which are identifiable and correctable,
\be
Y_1|\tilde{\psi}\rangle&=&i[a[|100\rangle+|111\rangle-|001\rangle-|010\rangle]\non\\
&&+b[-|011\rangle-|000\rangle+|110\rangle+|101\rangle]],\non\\
Y_2|\tilde{\psi}\rangle&=&i[a[|010\rangle-|001\rangle+|111\rangle-|100\rangle]\non\\
&&+b[-|101\rangle+|110\rangle-|000\rangle+|011\rangle]],\non\\
Y_3|\tilde{\psi}\rangle&=&i[a[|001\rangle-|010\rangle-|100\rangle+|111\rangle]\non\\
&&+b[-|110\rangle+|101\rangle+|011\rangle-|000\rangle]].\non
\ee
The measurement-free QEC protocol is shown in Fig.~(\ref{figure1}d).

\emph{Performance for the general Pauli noise.}---The general Pauli noise is represented by $\varepsilon_{0}$,
\be
\varepsilon_{0}=f_{0}\hat{\sigma}_{I}+p_{x}\hat{\sigma}_{x}+p_{y}\hat{\sigma}_{y}+p_{z}\hat{\sigma}_{z}.\non
\ee
Here, $f_{0}$ is the channel fidelity, and $p_{x},p_{y},p_{z}$ is the probability of the Pauli-X, Pauli-Y, Pauli-Z, respectively. There are four different schemes for the general Pauli noise when it is the physical noise, and only the codes in Eq.~(\ref{e1}) and Eq.~(\ref{e2}) are available. The effective channel can be represented by $\varepsilon_{1}$ after one-level QEC.

For the code in Eq.~(\ref{e1}), when correcting $X_{1},X_{2},X_{3}$,
\be
\label{e3}
f_{1}&=&f_{0}^{3}+3f_{0}^{2}p_{x}+3f_{0}p_{z}^{2}+6f_{0}p_{y}p_{z}+3p_{x}p_{z}^{2}\non\\
p1_{z}&=&p_{z}^{3}+3p_{z}^{2}p_{y}+3p_{z}f_{0}^{2}+6f_{0}p_{x}p_{z}+3p_{y}f_{0}^{2}\non\\
p1_{x}&=&p_{x}^{3}+3p_{x}^{2}f_{0}+3p_{x}p_{y}^{2}+6p_{x}p_{y}p_{z}+3f_{0}p_{y}^{2}\non\\
p1_{y}&=&p_{y}^{3}+3p_{y}^{2}p_{z}+3p_{y}p_{x}^{2}+6f_{0}p_{y}p_{x}+3p_{z}p_{x}^{2}.
\ee
When correcting $Y_{1},Y_{2},Y_{3}$, the results can be obtained by replacing the $p_{x}$ with $p_{y}$, and the $p_{y}$ with $p_{x}$ in Eq.~(\ref{e3}).

For the code in Eq.~(\ref{e2}), when correcting $Z_{1},Z_{2},Z_{3}$,
\be
\label{e4}
f_{1}&=&f_{0}^{3}+3f_{0}^{2}p_{z}+3f_{0}p_{x}^{2}+6f_{0}p_{y}p_{x}+3p_{z}p_{x}^{2}\non\\
p1_{z}&=&p_{z}^{3}+3p_{z}^{2}f_{0}+3p_{z}p_{y}^{2}+6p_{z}p_{x}p_{y}+3f_{0}p_{y}^{2}\non\\
p1_{x}&=&p_{x}^{3}+3p_{x}^{2}p_{y}+3p_{x}f_{0}^{2}+6p_{x}p_{z}f_{0}+3p_{y}f_{0}^{2}\non\\
p1_{y}&=&p_{y}^{3}+3p_{y}^{2}p_{x}+3p_{y}p_{z}^{2}+6p_{y}f_{0}p_{z}+3p_{x}p_{z}^{2}.
\ee
When correcting $Y_{1},Y_{2},Y_{3}$, the results can be obtained by replacing the $p_{z}$ with $p_{y}$, and the $p_{y}$ with $p_{z}$ in Eq.~(\ref{e4}).

Here, $f_{1}$ is the channel fidelity, and $p1_{x},p1_{y},p1_{z}$ is the probability of the Pauli-X, Pauli-Y, Pauli-Z, respectively. In general cases, $f_{0}\gg p_{x},p_{y},p_{z}$. So, based on the value of $p1_{z}$ in Eq.~(\ref{e3}) and $p1_{x}$ in Eq.~(\ref{e4}), the three-qubit codes have the ability to polarize the effective channel. The code in Eq.~(\ref{e1}) will polarize the effective channel to Pauli-Z, and the code in Eq.~(\ref{e2}) will polarize the effective channel to Pauli-X. This phenomenon makes the effective channel more correctable, and increases the possibility of realizing efficient QEC with the three-qubit codes.

\emph{Rules of concatenation.}---Based on the Eq.~(\ref{e3}) and Eq.~(\ref{e4}), $4^{8}$ different combinations with the four Pauli operators are correctable when concatenated the two codes in Eq.~(\ref{e1}) and Eq.~(\ref{e2}). Fortunately, the correctable errors contain all one-wight errors, which is the necessary condition for efficient QEC. Now, the core problem we need to solve becomes how to improve performance under the premise of efficient QEC. There are four rules we should follow. For any independent noise, the kraus operators can be written as,
\be
\varepsilon(\rho)=\sum_{i}E_{i}\rho E^{\dag}_{i}.\non
\ee
Then, $i$ always can be set as $4$. To determine the choice of the four QEC protocol in Fig.~(\ref{figure1}), we define the similarity between any independent noise and Identity, or Pauli-X, or Pauli-Z, or Pauli-Y,
\be
\label{e5}
S_{I}&=&\frac{1}{4}\sum^{4}_{i}(Tr[E_{i}.\hat{\sigma}_{I}])^{2},\non\\
S_{X}&=&\frac{1}{4}\sum^{4}_{i}(Tr[E_{i}.\hat{\sigma}_{x}])^{2},\non\\
S_{Z}&=&\frac{1}{4}\sum^{4}_{i}(Tr[E_{i}.\hat{\sigma}_{z}])^{2},\non\\
S_{Y}&=&|\frac{1}{4}\sum^{4}_{i}(Tr[E_{i}.\hat{\sigma}_{y}])^{2}|.
\ee
Because the operators $\hat{\sigma}_{I},\hat{\sigma}_{x},\hat{\sigma}_{z},\hat{\sigma}_{y}$ can be viewed as one complete set of basis vectors, so $S_{I}+S_{X}+S_{Z}+S_{Y}=1$. Now, we can classify all cases as $4$ categories through Eq.~(\ref{e5}), and applying QEC protocol in Fig.~(\ref{figure1}) accordingly,

\emph{1. $S_{Y}<S_{X}\geq S_{Z}$, applying QEC in Fig.~(\ref{figure1}a).}

\emph{2. $S_{Y}<S_{Z}>S_{X}$, applying QEC in Fig.~(\ref{figure1}b).}

\emph{3. $S_{Y}\geq S_{Z}\geq S_{X}$, applying QEC in Fig.~(\ref{figure1}c).} The code in Eq.~(\ref{e1}) has the stabilizer $\langle Z_1Z_2, Z_2Z_3\rangle$, higher effective channel fidelity can be obtained than applying Fig.~(\ref{figure1}d).

\emph{4. $S_{Y}\geq S_{X}>S_{Z}$, applying QEC in Fig.~(\ref{figure1}d).} The code in Eq.~(\ref{e2}) has the stabilizer $\langle X_1X_2, X_2X_3\rangle$, higher effective channel fidelity can be obtained than applying Fig.~(\ref{figure1}c).

However, after QEC applied, the effective channel also can be classified through Eq.~(\ref{e5}). Then, applying the next level QEC under the four rules.

\emph{The cost qubits, gate operations and fault tolerance threshold.}---For the concatenated level-$l$ QEC protocol with the three-qubit codes, the qubits needed can be obtained,
\be
\label{e6}
Q_{l}=3^{l}.
\ee
Meanwhile, the number of gate operations can be obtained,
\be
\label{e7}
N_{l}=2(n_{1}3^{l-1}+n_{2}3^{l-2}+n_{3}3^{l-3}\cdot\cdot\cdot+n_{l}3^{l-l}).
\ee
In the equation, $n_{1},n_{2},n_{3}\cdot\cdot\cdot n_{l}$ are the numbers of the gate operations in Fig.~(\ref{figure1}), which is $3$ when applied QEC in Fig.~(\ref{figure1}a), or $5$ when applied QEC in Fig.~(\ref{figure1}b), or $5$ when applied QEC in Fig.~(\ref{figure1}c), or $7$ when applied QEC in Fig.~(\ref{figure1}d). Here, we should notice that the decoding protocol must be in Fig.~(\ref{figure1}), but the encoding protocol can be simplified,
\be
\label{e8}
N_{dl}&=&n_{1}3^{l-1}+n_{2}3^{l-2}+n_{3}3^{l-3}\cdot\cdot\cdot+n_{l}3^{l-l},\non\\
N_{el}&=&n_{e1}3^{l-1}+n_{e2}3^{l-2}+n_{e3}3^{l-3}\cdot\cdot\cdot+n_{el}3^{l-l}.
\ee
In the equation, $n_{e1},n_{e2},n_{e3}\cdot\cdot\cdot n_{el}$ are the numbers of the gate operations in encoding, which is $2$ when applied QEC in Fig.~(\ref{figure1}a,c), or $4$ when applied QEC in Fig.~(\ref{figure1}b,d). And $3^{l-1}, 3^{l-2}, 3^{l-3}\cdot\cdot\cdot3^{l-l}$ is the number of encoding/decoding module.

Based on the Eq.~(\ref{e8}), the accuracy rate of the concatenated-$l$ QEC protocol can be defined,
\be
\label{e9}
R_{l}=r^{(n_{e1}+n_{e2}+n_{e3}\cdot\cdot\cdot+n_{el})+(n_{1}+n_{2}+n_{3}\cdot\cdot\cdot+n_{l})}.
\ee
Here, $r$ is the accuracy rate for every gate operation, and $r^{n_{el}}$ ($r^{n_{l}}$) is the accuracy rate of the $l$-th concatenation for encoding (decoding). Because the numbers $n_{e1},n_{e2},n_{e3}\cdot\cdot\cdot n_{el}$ and $n_{1},n_{2},n_{3}\cdot\cdot\cdot n_{l}$ are small, the accuracy rate of the protocol $R_{l}$ has a slow damping.

As the comparison, we choose the five-qubit code in ref.~\cite{Bennett}. For concatenated level-$l$ QEC protocol with the five-qubit code, the the number of qubits needed is $Q_{l}=5^{l}$. The number of gate operations can be obtained,
\be
\label{e10}
N_{dl}&=&n_{1}5^{l-1}+n_{2}5^{l-2}+n_{3}5^{l-3}\cdot\cdot\cdot+n_{l}5^{l-l},\non\\
N_{el}&=&n_{e1}5^{l-1}+n_{e2}5^{l-2}+n_{e3}5^{l-3}\cdot\cdot\cdot+n_{el}5^{l-l}.
\ee
In the equation, $n_{1}=n_{2}=n_{3}\cdot\cdot\cdot=n_{l}=22$ are the numbers of the gate operations for decoding, and $n_{e1}=n_{e2}=n_{e3}\cdot\cdot\cdot=n_{el}=15$ are the numbers of the gate operations for encoding. So, the accuracy rate of the concatenated-$l$ QEC protocol can be defined,
\be
\label{e11}
R_{l}&=&r^{(n_{e1}+n_{e2}+n_{e3}\cdot\cdot\cdot+n_{el})+(n_{1}+n_{2}+n_{3}\cdot\cdot\cdot+n_{l})}\non\\
&=&r^{15l+22l}.
\ee
As shown, the numbers $n_{e1}=n_{e2}=n_{e3}\cdot\cdot\cdot=n_{el}=15$ and $n_{1}=n_{2}=n_{3}\cdot\cdot\cdot=n_{l}=22$, the accuracy rate of the protocol with five-qubit code has a faster damping.

Now, based on the accuracy rate $R_{l}$ of the concatenated-$l$ QEC protocol and the effective channel fidelity $f_{l}$, the fidelity $F_{l}$ in reality can be obtained,
\be
\label{e12}
F_{l}=R_{l}\cdot f_{l}.
\ee
One QEC protocol that can work effectively in reality must ensure,
\be
\label{e13}
F_{l}>F_{l-1}, l\geq1.
\ee
Then, the QEC protocol can be one fault tolerance QEC protocol, which can be realized in the physical system. Through calculations, it is found that the concatenated QEC with the modules in Fig.~(\ref{figure1}) has great potential for realizing QEC in reality.

\emph{Examples.}---

\emph{1. Depolarizing channel with the initial fidelity $f_{0}=0.92$.}For the concatenated $4$ levels QEC protocol with three-qubit codes, and the accuracy rate $r=\sqrt{0.999}$ for every gate operation. Based on the rules of concatenation, we can obtain,
\be
\label{e14}
Q_{4}&=&3^{4}=81,\non\\
N_{d4}&=&5\times3^{4-1}+5\times3^{4-2}+5\times3^{4-3}+3\times3^{4-4}=198,\non\\
N_{e4}&=&2\times3^{4-1}+4\times3^{4-2}+4\times3^{4-3}+2\times3^{4-4}=104,\non\\
R_{4}&=&\sqrt{0.999}^{(2+4+4+2)+(5+5+5+3)}=0.985105,\non\\
F_{4}&=&0.985105\times0.960219=0.945917.
\ee
For the concatenated $3$ levels QEC protocol with five-qubit code, and the accuracy rate $r=\sqrt{0.999}$ for every gate operation. We can obtain,
\be
\label{e15}
Q_{3}&=&5^{3}=125,\non\\
N_{d3}&=&22\times5^{3-1}+22\times5^{3-2}+22\times5^{3-3}=682,\non\\
N_{e3}&=&15\times5^{3-1}+15\times5^{3-2}+15\times5^{3-3}=465,\non\\
R_{3}&=&\sqrt{0.999}^{(15+15+15)+(22+22+22)}=0.945986,\non\\
F_{3}&=&0.945986\times0.993991=0.940301.
\ee

\emph{2. Amplitude damping noise with the initial fidelity $f_{0}=0.9$.} For the concatenated $4$ levels QEC protocol with three-qubit codes, and the accuracy rate $r=\sqrt{0.999}$ for every gate operation. Based on the rules of concatenation, we can obtain,
\be
\label{e16}
Q_{4}&=&3^{4}=81,\non\\
N_{d4}&=&7\times3^{4-1}+3\times3^{4-2}+3\times3^{4-3}+5\times3^{4-4}=230,\non\\
N_{e4}&=&4\times3^{4-1}+2\times3^{4-2}+2\times3^{4-3}+4\times3^{4-4}=136,\non\\
R_{4}&=&\sqrt{0.999}^{(4+2+2+4)+(7+3+3+5)}=0.985105,\non\\
F_{4}&=&0.985105\times0.961634=0.94731.
\ee
For the concatenated $3$ levels QEC protocol with five-qubit code, and the accuracy rate $r=\sqrt{0.999}$ for every gate operation. We can obtain,
\be
\label{e17}
Q_{3}&=&5^{3}=125,\non\\
N_{d3}&=&22\times5^{3-1}+22\times5^{3-2}+22\times5^{3-3}=682,\non\\
N_{e3}&=&15\times5^{3-1}+15\times5^{3-2}+15\times5^{3-3}=465,\non\\
R_{3}&=&\sqrt{0.999}^{(15+15+15)+(22+22+22)}=0.945986,\non\\
F_{3}&=&0.945986\times0.975488=0.922798.
\ee

The concatenated level for five-qubit code can not be greater than $3$, because the accuracy rate of overall QEC protocol decays too fast to bring any improvement on the fidelity in reality. The results indicate that the concatenated three-qubit codes has greater advantage than the concatenated five-qubit code because of the higher accuracy rate. However, if we want to achieve $10^{-5}$ quantum error correction in reality, higher accuracy rate of the quantum gate operation is required (perhaps $r=1-10^{-6}$ and $f_{0}=0.99$). It should be noticed that $f_{l}\propto l$, $R_{l}\propto r$, and $R_{l}\propto \frac{1}{l}$, the max value of $F_{l}$ may not be necessarily at the maximum $l$.

\emph{Conclusions and outlook.}---In this work, the efficient QEC protocol against the general independent noise is constructed with the three-qubit QEC codes. The rules of concatenation are summarized according to the error-correcting capability of the codes. The codes not only play the role of correcting errors, but the role of polarizing the effective channel. The concatenated three-qubit codes has the potential to realize effective QEC in reality, and has great advantage than the concatenated five-qubit code. Not only less sources are needed, but better QEC performance is obtained.

In the realization of QEC, state preservation by repetitive error detection has been done in a superconducting quantum circuit~\cite{Kelly}, and estimation of consumption have been done in~\cite{Sohn}. Why the concatenated QEC protocol can be efficient, the reason may arise from the ability to change the logical error when applying inner QEC, and the outer QEC is good at suppressing it. Similar phenomena for other codes have been noted in~\cite{Nielsen,Huang1,Huang2}, but the realizations need more complicated quantum gate operations. As an example in realization, once the QEC with the three-qubit code had been realized as in~\cite{Chiaverini}. Based on the results in this work, we believe that QEC could be realized when a quantum gate operation is technically accurate enough.

\end{document}